\documentclass[%
 aip,
 utf8,
 cha, 
 reprint,%
 numerical,
]{revtex4-1}

\usepackage[english]{babel}
\usepackage[utf8]{inputenc}

\usepackage{graphicx}    
\usepackage{dcolumn}     
\usepackage{bm}          
\usepackage{titlesec}
\usepackage{color}       
\usepackage{ulem}        

\usepackage{subfigure}
\usepackage{enumitem}

\usepackage{relsize}    
\def\textsubscript#1{\ensuremath{_{\mbox{\textscale{.6}{#1}}}}}

\begin{document}


\title[Produtos e Materiais Didáticos]{\LARGE{Looking to the Blue Sky with Colored Patterns}\\ \small{Olhando o Céu Azul com Padrões Coloridos}}  

\author{Diogo Soga}
\email[E-mail:]{diogosp@usp.br}
\affiliation{Institute of Physics, University of Sao Paulo} 
\author{Daniel M. Faes}%
\affiliation{Institute of Astronomy, Geophysics and Atmospheric Sciences, University of Sao Paulo}
\author{Mikiya Muramatsu}%
\affiliation{Institute of Physics, University of Sao Paulo} 

\date{\today}

\begin{abstract}
This work presents a simple alternative technique to observe the polarization of the skylight on Earth. Using a birefringent material and a polarizer to look at the polarized light from the blue sky, it is possible to see a colored pattern that is associated to the birefringence of the material and the polarized light. Three different ways to polarize the light are also discussed in the context of the proposed experiment. 
\\
Este trabalho apresenta uma técnica alternativa para  visualizar os efeitos da luz polarizada do céu azul da Terra. Usando um material birrefringente e um polarizador para olhar o céu azul, aparece um padrão colorido que é associado a birrefringência do material e a luz polarizada. Também são discutidos três processos de polarização da luz no contexto deste experimento. \\
\textbf{Palavras-chave:} Céu azul, espalhamento de Rayleigh, birrefringência, polarização, ensino. \\

\end{abstract}

                                                            
\maketitle

\section{\label{sec:introducao} Introduction}

    Polarization of the light is an important matter because it is related to several current technological products, such as monitors, displays, and glasses with liquid crystals; 3D movies projectors; sunglasses and lasers. 

    Physics textbooks~\cite{young2008, serway1982,halliday1992} describe the four physical processes that polarize the light: reflection~\cite{bedran1997}, selective absorption, scattering and double refraction (birefringence). This work discusses the last three processes. 
		
\begin{figure}[hbtp]
    \begin{center}
        \includegraphics*[width=8.0 cm]{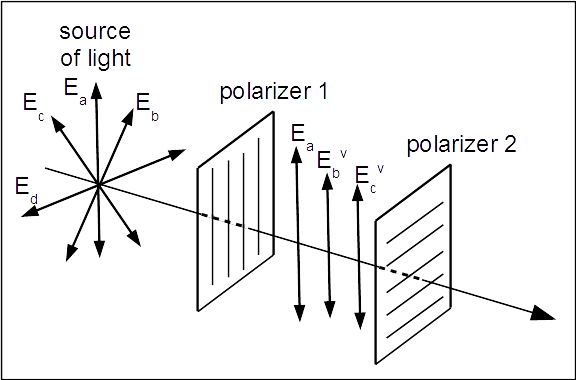}
      \caption{Process of polarization by selective absorption.}
        \label{fig:luz}
    \end{center}
\end{figure}

    The selective absorption occurs in materials that transmit waves whose electric fields vibrate in a plane parallel to a certain direction and that absorb waves whose electric fields vibrate in all other directions.
    Here the materials with this characteristic are named  ``polarizers''. 
    This process is illustrated in figure \ref{fig:luz}. Considering a non-polarized source of light (the left side), the rays of light travel to right side passing through the polarizer 1, which polarization direction is vertical. Only the rays \normalsize E\textsubscript{a}, E\textsubscript{b}\textsuperscript{v},  and E\textsubscript{c}\textsuperscript{v} are transmitted and the others are absorbed. The rays \normalsize E\textsubscript{b}\textsuperscript{v}  and E\textsubscript{c}\textsuperscript{v} are the vertical components of the rays \normalsize E\textsubscript{b} and E\textsubscript{c}, respectively. If there is a second polarizer (polarizer 2) and it is oriented orthogonally to the first polarizer, then no light pass through it. If the polarization direction of the second polarizer is in other orientation, partial or total light intensity could pass through it. 

\begin{figure}[hbtp]
    \begin{center}
        \mbox{
            \subfigure[Parallel directions.]{\includegraphics*[width=3.5 cm]{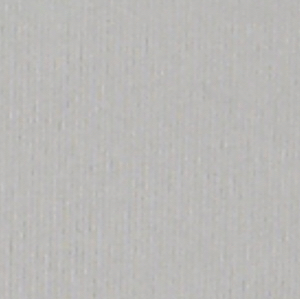}\label{fig:pol_paralelos}} 
            
            \subfigure[Orthogonal directions.]{\includegraphics*[width=3.5 cm]{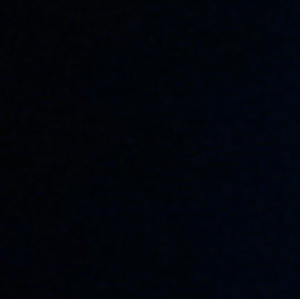}\label{fig:pol_cruzados}}
    }
    \caption{Images of the polarized light from a liquid crystal when observed after passing through a polarizer.}
    \label{fig:polarizador}
    \end{center}
\end{figure}

    The figure \ref{fig:polarizador} shows two images from a liquid crystal monitor, which emits polarized light, when observed after passing through a polarizer. In the first case (fig.~\ref{fig:pol_paralelos}), the polarization direction of the polarizer is parallel to the emitted light, so that the transmitted intensity is maximum. The color is not white because the polarizer absorbs a small part of the light. In the figure~\ref{fig:pol_cruzados}, the direction of the polarizer is orthogonal to the light, corresponding to the minimum light intensity.

    The processes of polarization by light scattering and double refraction (birefringence) will be discussed in the next sections.

    This work presents an alternative technique to see the effect of polarized light using a birefringent material. This type of material could modify the orientation of the polarization direction after passing through it.

\section{\label{sec:espalhamento} Light Scattering}

\begin{figure}[hbtp]
    \begin{center}
        \includegraphics*[width=8.2 cm]{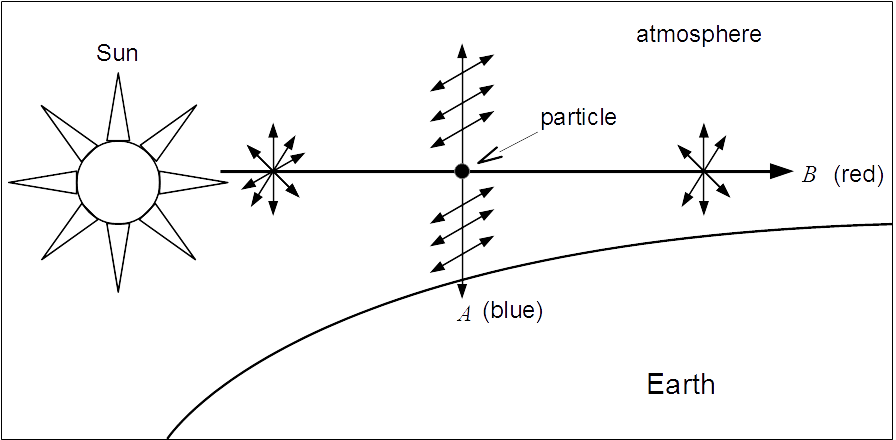}
            
    \caption{The process of Rayleigh scattering of the blue sky of Earth. \textit{A} and \textit{B} are points of observation.~\cite{young2008}}
    \label{fig:ceu_azul}
    \end{center}
\end{figure}

    The blue color of the sky \cite{serway1982, halliday1992, sakurada2002, sharma2003, young2008} of planet Earth is explained by the theory of Rayleigh scattering. The white light from the Sun (fig.~\ref{fig:ceu_azul}) is scattered by molecules or particles in the atmosphere of the planet, in a process that depends on the wavelength of the light ($\lambda$), or more precisely, is proportional to the inverse of $\lambda^4$. The scattering of the light by particles is more efficient at short wavelengths than at longer ones. The interval of the visible light spectrum~\cite{hecht1998} varies from 390\,nm until 700\,nm; the visible spectrum is divided into six colors: violet (390\,nm), blue, green, yellow, orange, and red (700\,nm). The sky does not appear violet color because the human eye~\cite{bansal2005} has low sensitivity to this color. The next color in the spectrum is the blue, where the human eye has a better sensitivity. This is the reason why the sky is blue. 
		
Figure~\ref{fig:ceu_azul} illustrates how these effects are visualized. At the point \textit{A} (fig.~\ref{fig:ceu_azul}), a person looking up sees the scattered light of the Sun or the sky in blue. A person in point \textit{B}, which corresponds to looking at the Sun at sunset, sees the sky in red since this color is the least scattered.

     The Rayleigh theory describes that part of the scattered light is polarized. It is possible to see the effect of this polarization using a polarizer, rotating it so that the intensity of transmitted light varies. However it is not a simple process because according to Lynch~\cite{lynch1995} the light is polarized at the maximum of 85\%, and to sense small variations of intensity can be hard.

\begin{figure}[hbtp]
    \begin{center}
        \includegraphics*[width=8.00 cm]{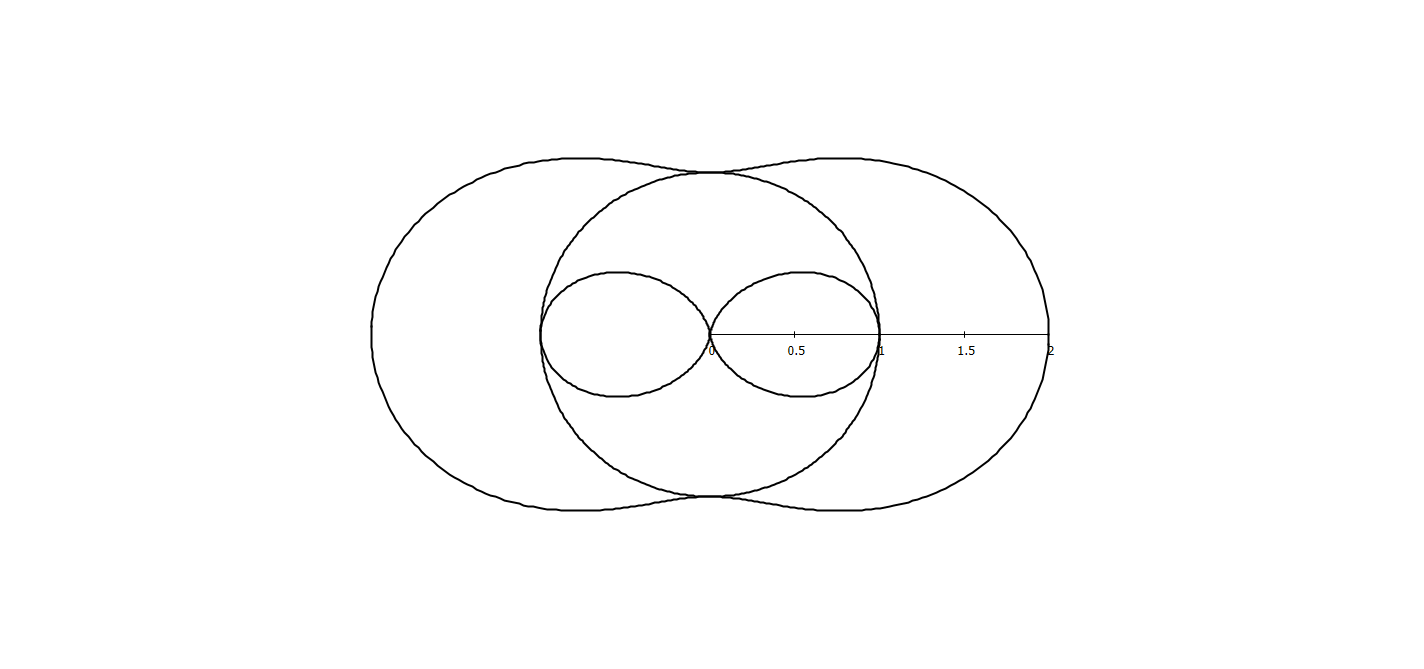}
        \caption{The intensity distribution of the scattered radiation by a particle. This is a polar plot illustrates the sky polarization by considering the incoming light of the Sun from the left and the scattering particle in the atmosphere at the center~\cite{lynch1995}.}
        \label{fig:rayleigh}
    \end{center}
\end{figure}

    A simple approach to the Rayleigh theory is to consider the particle as a sphere with radius $a$, that is much smaller~\cite{seinfeld2006} than $\lambda$ (concerning to light in the visible range of the spectrum, particles of diameter $\le$ 0.1\,$\mu m$ ), and that irradiates as an electrical dipole. The intensity of the scattered light ($I(\theta)$) from the particle is described as\cite{sakurada2002}:
\begin{eqnarray}
 I(\theta)  = \frac{8 \pi^4 a^6}{R^2 \lambda^4} \left|\frac{m^2 - 1}{m^2 + 2}\right|^2 (1 + \cos^2 \theta), \label{eq:rayleigh}
\end{eqnarray}
where $\theta$ is the scattering angle, $R$ is the distance of the dipole, $m$ is the 
ratio of two values of refraction index, of the air and the particle.
 The equation \ref{eq:rayleigh} could be rewritten as:
\begin{eqnarray}
     I(\theta)  = I_{1} I_{2},
\end{eqnarray}
where:
\begin{eqnarray}
   I_1  = \frac{8 \pi^4 a^6}{R^2 \lambda^4} \left|\frac{m^2 - 1}{m^2 + 2}\right|^2,
\end{eqnarray}
and
\begin{eqnarray}
     I_2  = (1 + \cos^2 \theta).
\end{eqnarray}
The term $I_{1}$ has no dependence with $\theta$; then the scattered intensity is uniform for all directions (fig. \ref{fig:rayleigh}). However the second term ($I_{2}$) has dependence with $\cos (\theta)$, what corresponds to two directions without scattering (at $\theta =\pm90^o$), or the perpendicular directions of the incident light. Then the scattered light $I(\theta)$ is polarized in these directions. 

\begin{figure}[hbtp]
    \begin{center}
        \mbox{
        \subfigure[Degree = $0^o$.]{\includegraphics*[width=3.5 cm]{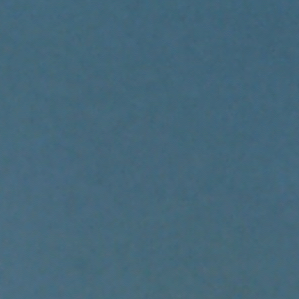}\label{fig:ceu_1_g000}} 
            
            \subfigure[Degree = $45^o$.]{\includegraphics*[width=3.5 cm]{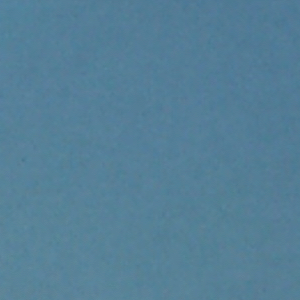}\label{fig:ceu_1_g045}}
    }

     \mbox{
        \subfigure[Degree = $90^o$.]{\includegraphics*[width=3.5 cm]{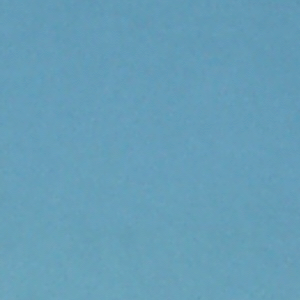}\label{fig:ceu_1_g090}} 
            
            \subfigure[Degree = $135^o$.]{\includegraphics*[width=3.5 cm]{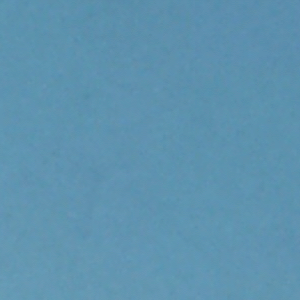}\label{fig:ceu_1_g135}}
}

     \mbox{
        \subfigure[Degree = $180^o$.]{\includegraphics*[width=3.5 cm]{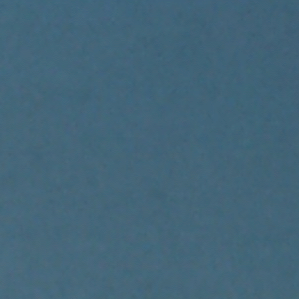}\label{fig:ceu_1_g180}} 
}
        \caption{The polarized light from the blue sky, analysed by a polarizer that rotated. Author: D. Soga.}
        \label{fig:ceu_polarized}
    \end{center}
\end{figure}

The figure~\ref{fig:ceu_polarized} presents five images of the polarized light from the blue sky on a sunny day. These images were captured through a polarizer (selective absorption type) in front of a camera and the polarizer was rotated as indicated. The first image (fig.~\ref{fig:ceu_1_g000}) corresponds to the minimum intensity of light, then the direction of polarization of the polarizer is orthogonal to the direction of polarization of the blue sky. The intensity increases until rotation of $90^o$ and then decreases until rotation of $180^o$ (fig.~\ref{fig:ceu_1_g180}), which intensity is the same of the original orientation ($0^o$; fig.~\ref{fig:ceu_1_g000}).

    Under ideal conditions of a 100\% polarized light with a perfect polarizer (complete absorption outside the polarization orientation), and perfect orthogonal orientation between them, then no light is transmitted. The figure \ref{fig:ceu_1_g000} shows the minimum intensity, where part of the light is transmitted. The image at figure \ref{fig:ceu_1_g180} is similar to the image in figure \ref{fig:ceu_1_g000} after a rotation of 180$^o$. The ideal condition will not be present in such observations because the polarization of the sky is never complete. However the effects of polarized light can be easily observed using birefringent materials. 


\section{\label{sec:birrefringentes} Birefringence}

The birefringence is a characteristic present in some materials. Since it is an effect dependent on the different light propagation inside the material, they are also called as anisotropic materials. This class of materials~\cite{hecht1998} comprises the liquid crystals and some mineral crystals like the calcite.

\begin{figure}[hbtp]
    \begin{center}
        \includegraphics*[width=8.0 cm]{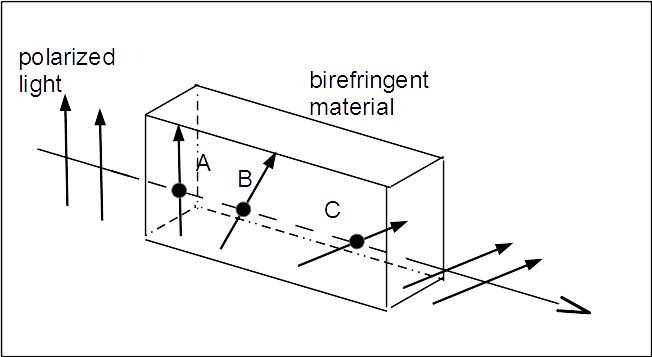}
    \caption{The polarized light changes its polarization direction after passing through a birefringent material.}
    \label{fig:atividade}
    \end{center}
\end{figure}

Some of the birefringent materials can rotate the polarization direction of the light. In figure~\ref{fig:atividade} the vector of polarized light is vertical, and after passing through the material, its direction is horizontal. Inside the material, the vector direction changes at each point. The rotation is a function of wavelength too.

If a birefringent material is put between a source of completely polarized white light and an orthogonal polarizer, part of the light would be necessarily transmitted, since the birefringent material would rotate the polarization direction, resulting a small component to be parallel to the polarizer. Then the transmitted light after the polarizer is not white, but a colored light. Each point of the material could rotate the light vector by a different value, then a colorful pattern appears. This behavior is used in the proposed experiment.

Some materials are intrinsically isotropic, but others can change their structure under a load condition, becoming anisotropic and then displaying birefringence. These materials, under illumination of white light and observed through a polarizer, present colored patterns, which are related to the stress distribution inside the material.



\section{\label{sec:materiais} Materials and Methods}

The experiment is described in this section. First, the materials with birefringence are presented. Following, the experimental procedure.

As a polarizer for the experiment, it can be used a polarizer filter of photographic cameras or a polarized sunglass lens,  or any other polarizer of the selective absorption type.

\subsection{\label{ssec:material} Plastic materials}

The birefringent materials employed were transparent pieces made of plastic: a protractor and a plate (from a box of CD/DVD). The structure of is kind of material can be sensible to stress. To verify this, polarized white light and a polarizer should be used. 

The following procedure was performed: these materials were illuminated with the polarized white light (e.g., from a liquid crystal monitor; see appendix \ref{ssec:ligth}), and the polarizer was placed after it. Then, the transmitted light was registered with a digital camera.

\begin{figure}[htb]
    \begin{center}
        \mbox{
        \subfigure[Without a polarizer.]{\includegraphics*[width=4.00 cm]{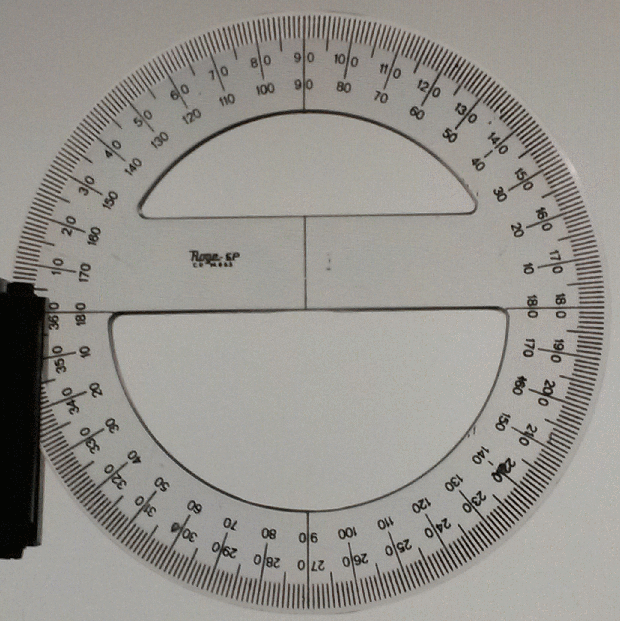}\label{fig:transferidor_claro}} 
            
            \subfigure[With a polarizer.]{\includegraphics*[width=3.97 cm]{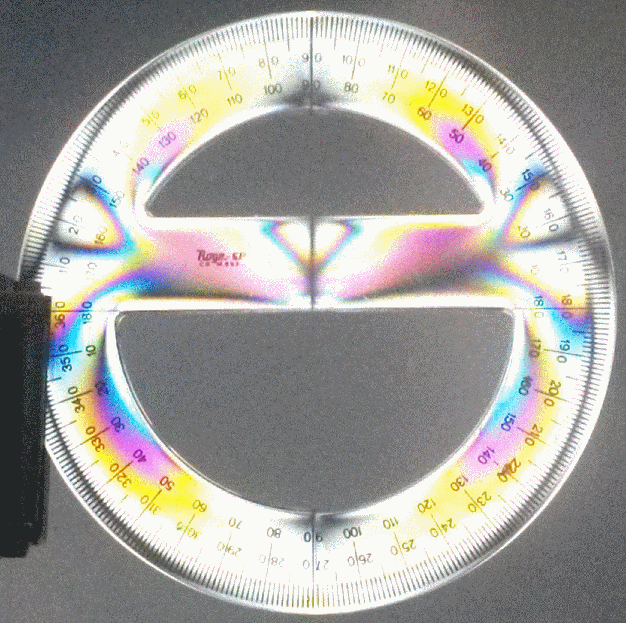}\label{fig:transferidor_escuro}}
         }
        \caption{A transparent protractor. Author: D. Soga.}
        \label{fig:transferidor_monitor}
    \end{center}
\end{figure}

Figure~\ref{fig:transferidor_claro} shows an image of a transparent protractor without the polarizer. It is possible to see the white light from the monitor and the scale. In figure~\ref{fig:transferidor_escuro}, the same passing through a polarizer. A colored pattern appears. This pattern is related to the distribution of stress inside the protractor, modifying the direction of polarization of the light from the monitor. The pattern is not present in the entire extension, and its shape is not uniform, but it appears that the right side is symmetric to the corresponding left side. This piece was used to do the majority of the experiments because it is easy to focus on the scale.

\begin{figure}[htb]
    \begin{center}
        \mbox{
        \subfigure{\includegraphics*[width=4.00 cm]{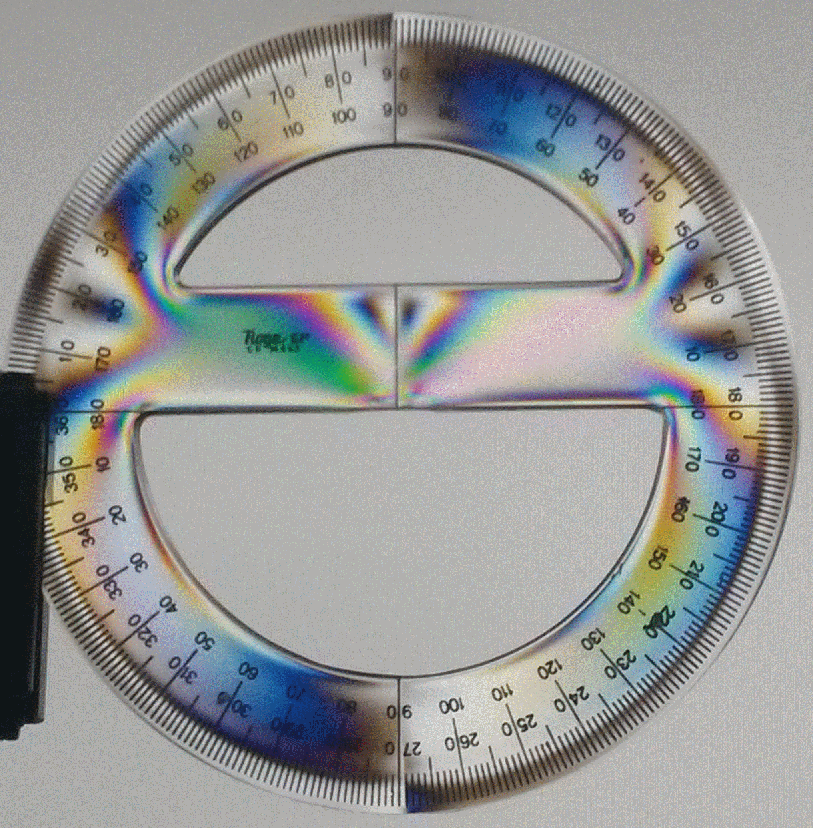}} 
    }
        \caption{A transparent protractor with the polarization direction of  the polarizer aligned to polarization of the incident light. Author: D. Soga.}
        \label{fig:transferidor_monitor_alinhado}
    \end{center}
\end{figure}

This technique does not depend on whether the direction of the polarizer is orthogonal to the direction of light polarization. In figure~\ref{fig:transferidor_monitor_alinhado} the colored pattern appears with the polarizer direction parallel to light polarization. In this case, the transmitted light is rotated, part of it is absorbed by the polarizer and a colorful pattern also appears. It is interesting to note that the pattern at the left side looks complementary to the pattern to the right side, and the colored fringes appear in places of one half size where does not appear in another half. This results in a different pattern of the one in figure~\ref{fig:transferidor_escuro}.

\begin{figure}[htpb]
    \begin{center}
        \mbox{
        \subfigure[Without a polarizer.]{\includegraphics*[width=4.00 cm]{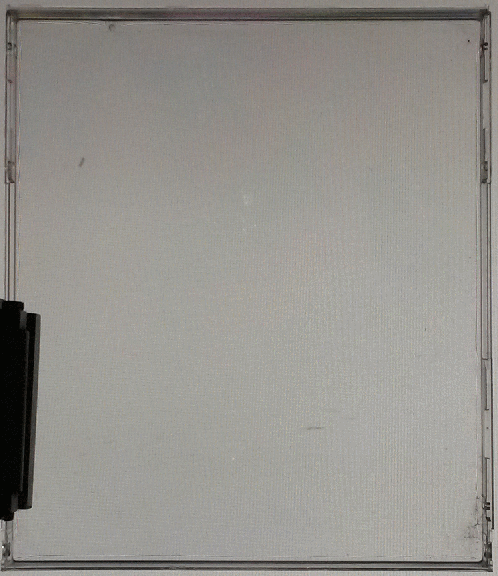}\label{fig:capa_claro}} 
            
            \subfigure[With a polarizer.]{\includegraphics*[width=3.95 cm]{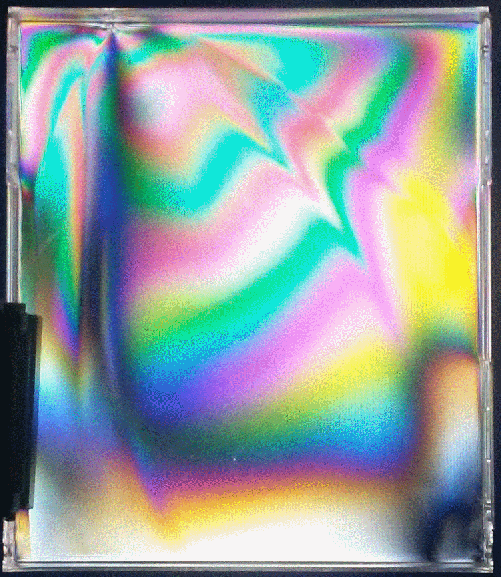}\label{fig:capa_escuro}}
         }
        \caption{A transparent plate from a box of CD/DVD. Author: D. Soga.}
        \label{fig:capa_monitor}
    \end{center}
\end{figure}

Figure~\ref{fig:capa_claro} shows an image of a transparent plate, where it is possible to see the white light from the monitor. If a polarizer is put between the material and the digital camera (fig.~\ref{fig:capa_escuro}), where the direction of the polarizer is orthogonal to the polarized light of the monitor, a colorful pattern appears, covering almost the entire piece. Its shape is different from the one in figure~\ref{fig:transferidor_escuro}. 
At the top of the piece, there is a small defect where are the most colored fringes. No symmetry is presented in this pattern.

\subsection{\label{ssec:experimento} The experiment}

The experiment consists of illuminating a birefringent material with the blue skylight, and adding a polarizer between the piece and the observer's eye. We discuss below images registered in August (winter season) in S\~ao Paulo city (Brazil). The experiment were done in several days, at different hours and in two different places. The experiment also experienced weather changes several times during the period.

We emphasize here as a precautionary measure to never look directly to the Sun.

\section{\label{sec:resultados} Results}

The analysis was done using only the half top of the piece because it was handled in the other half. The majority of the images were captured in open air.
    
\begin{figure}[htpb]
    \begin{center}
        \mbox{
        \subfigure[Direction north.]{\includegraphics*[width=4.0 cm]{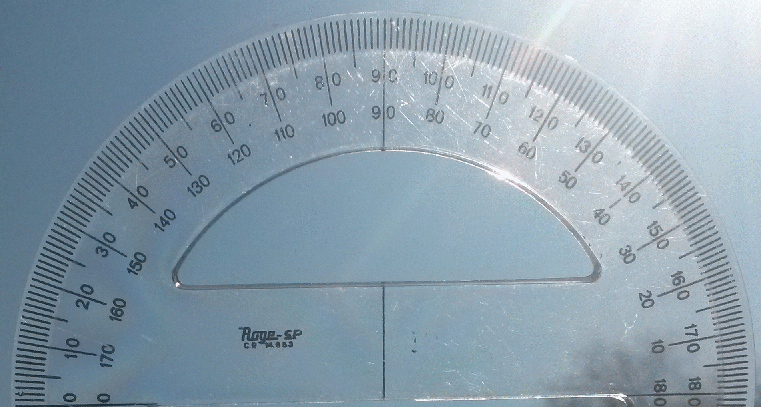}\label{fig:transfer_campo_n_170809}} 
            
            \subfigure[Direction south.]{\includegraphics*[width=4.0 cm]{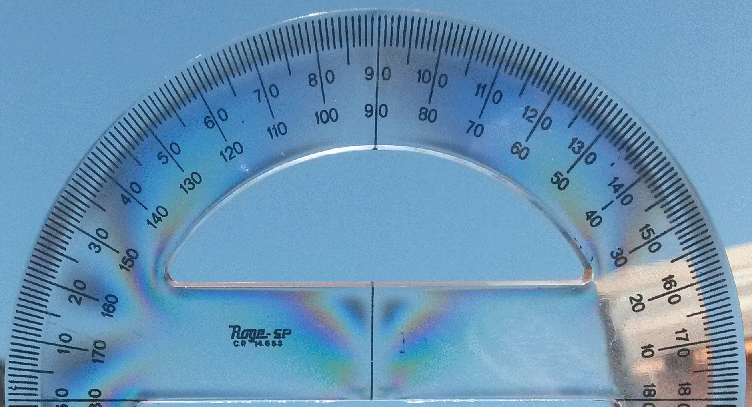}\label{fig:transfer_campo_s_170809}}
    }

     \mbox{
        \subfigure[Direction east.]{\includegraphics*[width=4.0 cm]{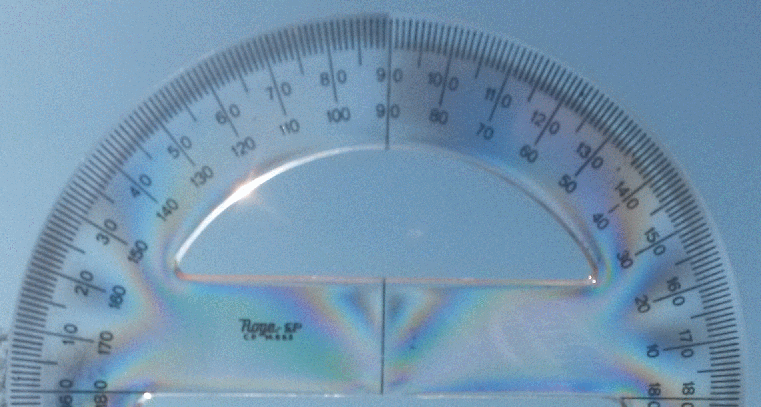}\label{fig:transfer_campo_l_170809}} 
            
            \subfigure[Direction west.]{\includegraphics*[width=4.0 cm]{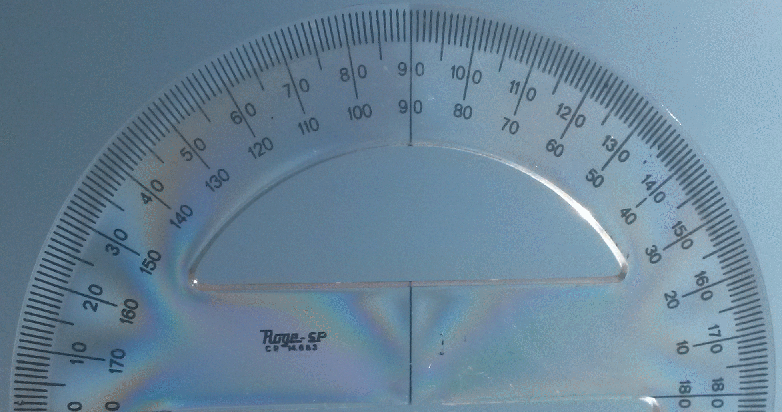}\label{fig:transfer_campo_o_170809}}
}

     \mbox{
        \subfigure[Direction zenith.]{\includegraphics*[width=4.0 cm]{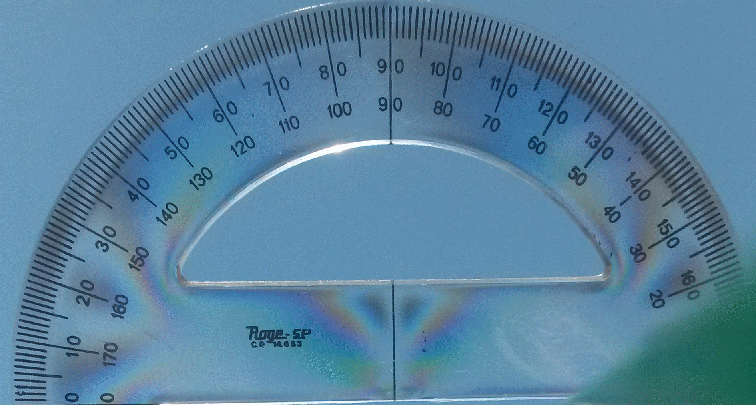}\label{fig:transfer_campo_az_170809}} 
}
        \caption{Images of a protractor illuminated by blue sky in a sunny day for four directions and the zenith. Author: D. Soga.}
        \label{fig:transferidor_campo_170809}
    \end{center}
\end{figure}

Figure~\ref{fig:transferidor_campo_170809} presents five images of the protractor (fig.~\ref{fig:transferidor_monitor_alinhado}) illuminated by the blue sky in sunny weather -- the four cardinal directions and the zenith (highest point of the sky). On one side of the piece, the colored pattern appears more prominently. Figure~\ref{fig:transfer_campo_n_170809} shows the image at the north direction, the colored pattern does not appear because the low polarization level since the Sun is near the visual field of the camera. However the colored pattern appears in figures~\ref{fig:transfer_campo_s_170809}, \ref{fig:transfer_campo_l_170809}, \ref{fig:transfer_campo_o_170809} and \ref{fig:transfer_campo_az_170809}, with no significant difference between them. In the south direction and in the zenith the pattern presented the best contrast. The shape of the colored pattern is similar to the one present in figure~\ref{fig:transferidor_monitor_alinhado}. From the background of each picture is possible to see the sky on a sunny day.

\begin{figure}[htpb]
    \begin{center}
        \mbox{
        \subfigure[Direction north.]{\includegraphics*[width=4.0 cm]{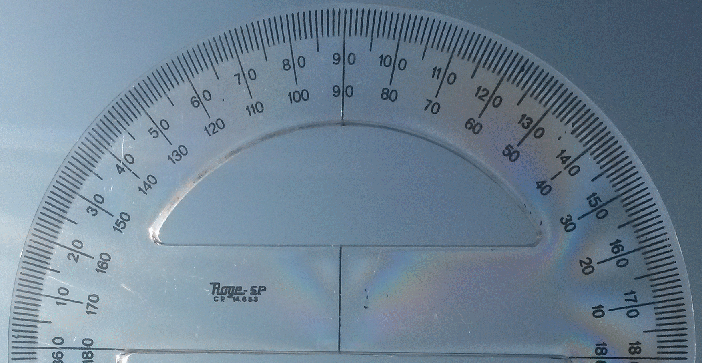}\label{fig:transfer_campo_n_170811}} 
            
            \subfigure[Direction south.]{\includegraphics*[width=4.0 cm]{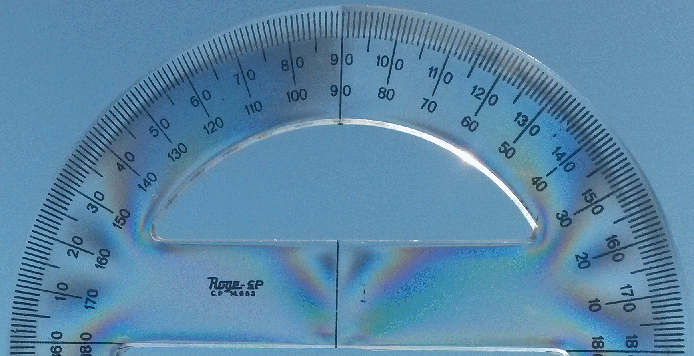}\label{fig:transfer_campo_s_170811}}
    }

     \mbox{
        \subfigure[Direction east.]{\includegraphics*[width=4.0 cm]{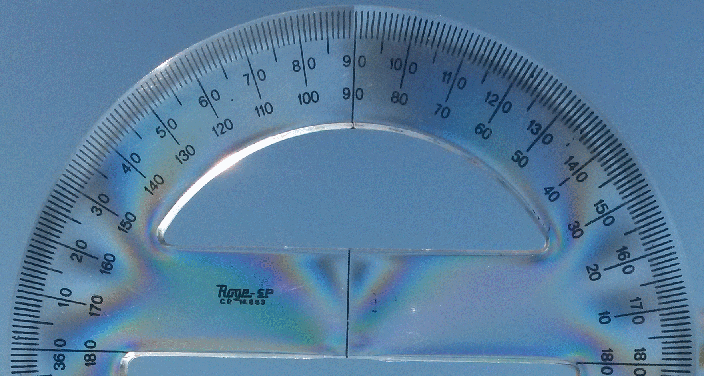}\label{fig:transfer_campo_l_170811}} 
            
            \subfigure[Direction west.]{\includegraphics*[width=4.0 cm]{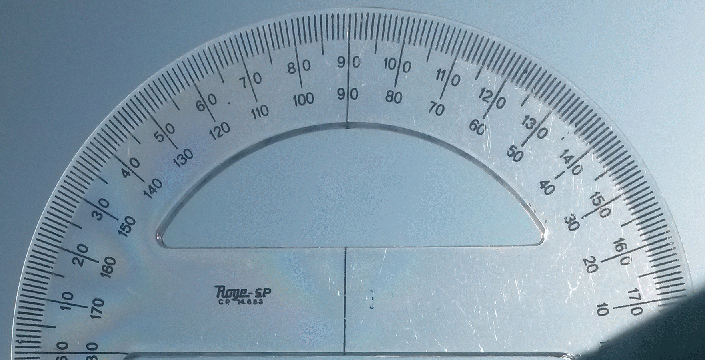}\label{fig:transfer_campo_o_170811}}
}

     \mbox{
        \subfigure[Direction zenith.]{\includegraphics*[width=4.0 cm]{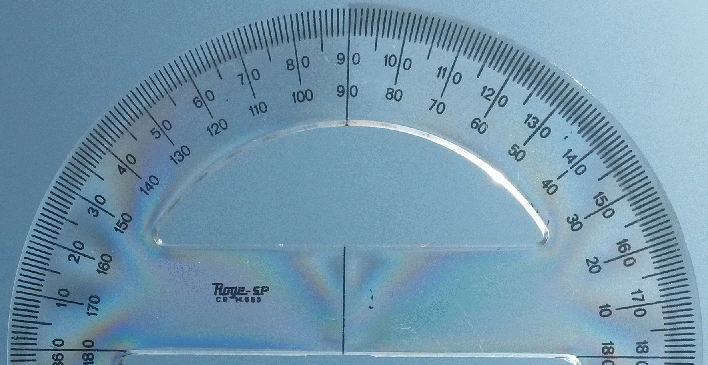}\label{fig:transfer_campo_az_170811}} 
}
        \caption{Additional images of a protractor illuminated by blue sky in a sunny day for four directions and zenith. Author: D. Soga.}
        \label{fig:transferidor_campo_170811}
    \end{center}
\end{figure}

Figure~\ref{fig:transferidor_campo_170811} presents additional five images of the protractor (fig.~\ref{fig:transferidor_monitor_alinhado}) illuminated by the blue sky in sunny weather, for the four directions and the zenith. In figure~\ref{fig:transfer_campo_n_170811} to north direction, the colored pattern appears only on the right side, because the Sun is on the left side of the image. The same occurs in figure~\ref{fig:transfer_campo_o_170811}, but in this case the Sun is on the right side, and the pattern appears on the left side. The colored pattern also appears in figures~\ref{fig:transfer_campo_s_170811}, \ref{fig:transfer_campo_l_170811} and \ref{fig:transfer_campo_az_170809}. Again the shape of the pattern is similar that one in figure~\ref{fig:transferidor_monitor_alinhado}.  

\begin{figure}[htpb]
    \begin{center}
        \mbox{
        \subfigure{\includegraphics*[width=4.0 cm]{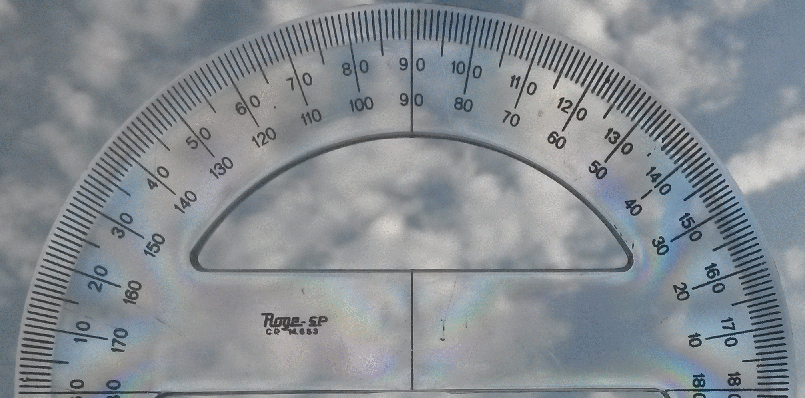}\label{fig:transfer_campo_nv_170813_a}} 
            
            \subfigure{\includegraphics*[width=4.0 cm]{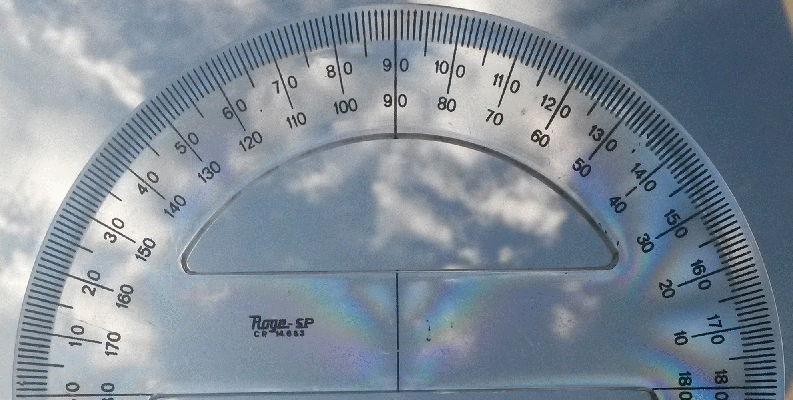}\label{fig:transfer_campo_nv_170813_b}}
    }

     \caption{The protractor illuminated by blue sky with the presence of clouds in the sky. Author: D. Soga.}
        \label{fig:transferidor_nuvens_170813}
    \end{center}
\end{figure}

\begin{figure}[htpb]
    \begin{center}
        \mbox{
        \subfigure[Direction north-west.]{\includegraphics*[width=4.0 cm]{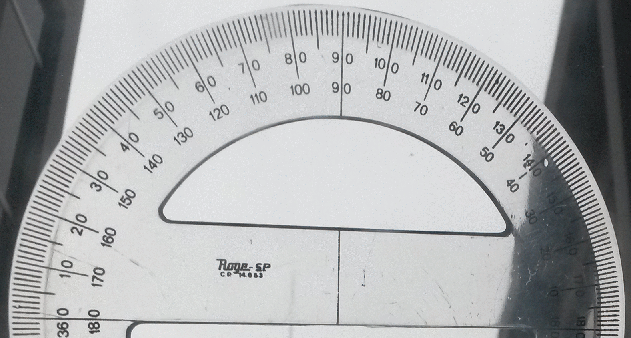}\label{fig:transferidor_campo_nublado_no}} 
            
            \subfigure[Direction north.]{\includegraphics*[width=4.0 cm]{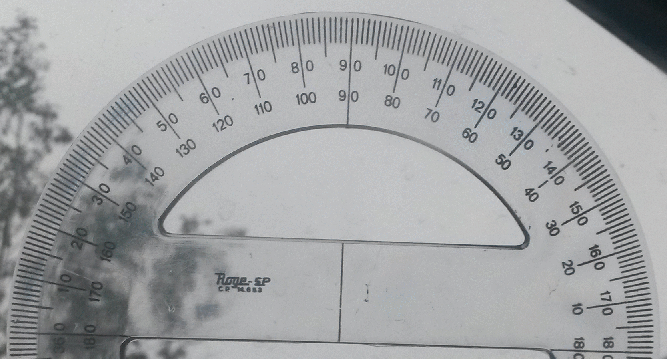}\label{fig:transferidor_campo_nublado_n}}
    }
     \caption{Images of the protractor illuminated by sky covered by clouds. Author: D. Soga.}
        \label{fig:transferidor_nublado}
    \end{center}
\end{figure}

The presence of few clouds in blue sky does not compromise the observations. The clouds disturb but do not destroy the colored pattern in both figures~\ref{fig:transfer_campo_nv_170813_a} and \ref{fig:transfer_campo_nv_170813_b}. However, when the sky is completely covered by clouds, the colored pattern does not appear, as one can see in the images in figure~\ref{fig:transferidor_nublado}.

\begin{figure}[htpb]
    \begin{center}
        \mbox{
        \subfigure[Rotation $30^0$.]{\includegraphics*[width=3.5 cm]{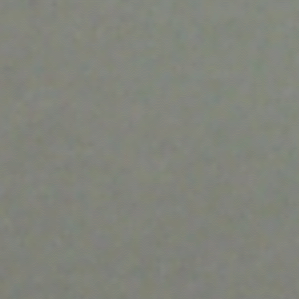}\label{fig:nublado_g030}} 
            
            \subfigure[Rotation $60^0$.]{\includegraphics*[width=3.5 cm]{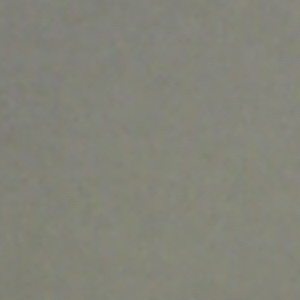}\label{fig:nublado_g060}}
    }

     \mbox{
        \subfigure[Rotation $90^0$.]{\includegraphics*[width=3.5 cm]{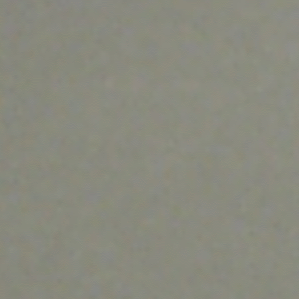}\label{fig:nublado_g090}} 
            
            \subfigure[Rotation $135^0$.]{\includegraphics*[width=3.5 cm]{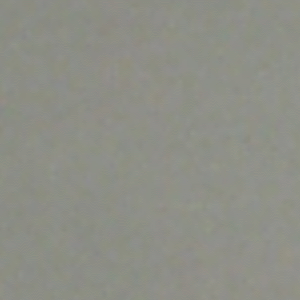}\label{fig:nublado_g135}}
}

        \caption{The skylight in a cloudy day through a polarizer at different orientations. Author: D. Soga.}
        \label{fig:nublado}
    \end{center}
\end{figure}

The polarized skylight was also tested using only the polarizer. The figure~\ref{fig:nublado} presents four images on a cloudy day for different polarized orientations. The intensity does not change considerably between each of them, indicating that the light has a low polarization level.

\begin{figure}[htpb]
    \begin{center}
        \mbox{
        \subfigure{\includegraphics*[width=7.0 cm]{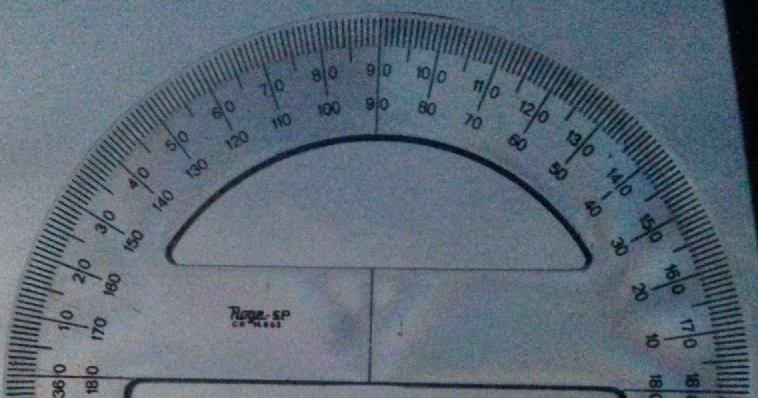}\label{fig:transfer_campo_nt_170813_a}} 
				}
				\mbox{
            
            \subfigure{\includegraphics*[width=7.0 cm]{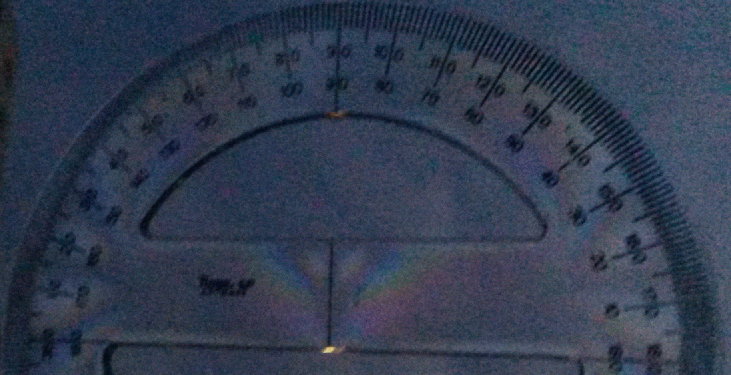}\label{fig:transfer_campo_nt_170813_b}}
         }

     \caption{The protractor illuminated by blue sky in the late afternoon. Author: D. Soga.}
        \label{fig:transferidor_noite_170813}
    \end{center}
\end{figure}

Another effect observed is that in the late afternoon the intensity of the polarized light decays, but the colored pattern can still be seen (fig.~\ref{fig:transferidor_noite_170813}).

\begin{figure}[htpb]
    \begin{center}
        \mbox{
        \subfigure[East]{\includegraphics*[width=4.00 cm]{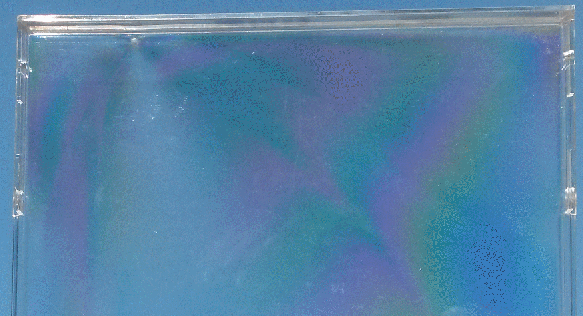}\label{fig:placa_l}} 
        \subfigure[West]{\includegraphics*[width=4.00 cm]{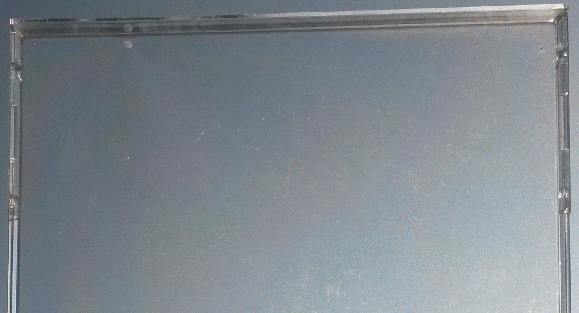}\label{fig:placa_o}} 
    }
        \mbox{
        \subfigure[North]{\includegraphics*[width=4.00 cm]{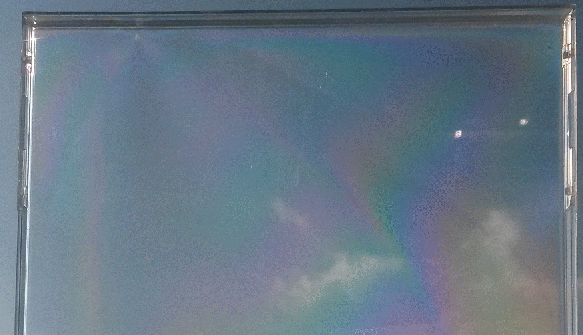}\label{fig:placa_n}} 
        \subfigure[South]{\includegraphics*[width=4.00 cm]{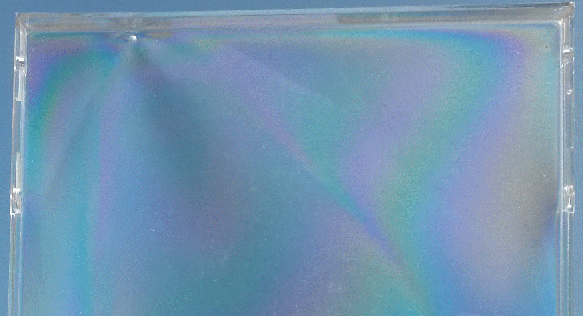}\label{fig:placa_s}} 
    }
        \mbox{
        \subfigure[Zenith]{\includegraphics*[width=4.00 cm]{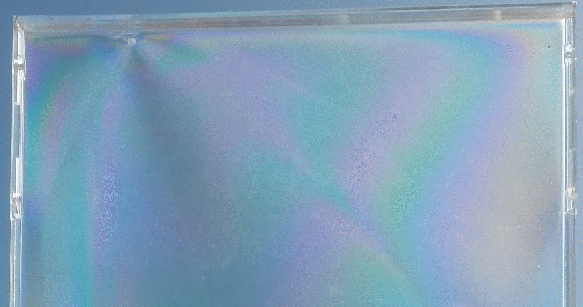}\label{fig:placa_az}} 
    }
        \caption{The plate illuminate by the blue sky, in a sunny day. Author: D. Soga.}
        \label{fig:capa_ceu}
    \end{center}
\end{figure}

    Figure~\ref{fig:capa_ceu} presents five images of the plastic plate, being illuminated by the blue sky on a sunny day. The colored pattern appears in all figures, except in figure~\ref{fig:placa_o}. The pattern is similar the one in figure~\ref{fig:capa_escuro}: the shape is the same, but the colors change. This indicates that there is a difference in the polarization in different directions, what could understand in terms of the distribution of scattered light shown in figure~\ref{fig:rayleigh}.
    
    This simple technique presents an alternative way to see the effect of the polarized light of the blue sky. The use of birefringent materials is an interesting tool to learn the concepts that light is a wave and that can be polarized. Moreover, that the white light is composed of multiple wavelengths. The colored pattern of these materials helps the students to realize that light from the blue sky is polarized. This activity could wake up the desire and motivate the students to conduct studies and analyzes of materials in a more carefully way.

\section{\label{sec:conclusao} Conclusion}

    This work presented an alternative technique to visualize the effects of the polarized light from the blue sky through the colored pattern that emerges from the birefringence of a transparent material. To see the colored pattern instead of a simple changing of the intensity of the light stimulates the students to understand polarization and the nature of the white light. This technique is sensible to different levels of polarized light, then the sky observations can be performed at different day hours, and even in the presence of a few clouds. Simple and cheap materials were used, such as plastic protracts and  plastic plates.
    
    Also, three processes of light polarization were discussed: the selective absorption, the scattering process, and the birefringence.

\section{\label{sec:agrado} Acknowledgment}
     DMF acknowledges support from FAPESP grant 2016/16844-1.


\appendix




\section{\label{ssec:ligth} Polarized White Light Source}

There are some simple polarized white light sources. One can be obtained using a liquid crystal monitor of computers. The emitted light is polarized, like as the one that can obtain using a white page of a text processor, like MS-Word or OpenOffice Writer, or a slide presentation software, like MS-Powerpoint or OpenOffice Impress. But liquid crystals displays with the touch-screen technology do not present a perfect polarized light.

A second option is to use a flashlight and a polarizer (fig. \ref{fig:luz}), since the emitted light by flashlight is not polarized. If it is a flashlight with more than one LED, it is necessary to use a thin white paper in the light beam to create an uniform illumination.

The third option is to use a white light bulb lamp and a polarizer.





\end{document}